\documentclass{nature}

\usepackage{epsfig}

\def\citep{\cite}
\def\citet{\cite}
\def\gtrsim{\mathrel{\hbox{\rlap{\hbox{\lower4pt\hbox{$\sim$}}}\hbox{$>$}}}}
\def\lesssim{\mathrel{\hbox{\rlap{\hbox{\lower4pt\hbox{$\sim$}}}\hbox{$<$}}}}

\def\degr{\hbox{$^\circ$}}
\def\arcmin{\hbox{$^\prime$}}

\def\degree{$^\circ$}

\def\arcs#1{$#1''$}
\def\arcsa#1#2{$#1^{\prime\prime}_{^\textrm{.}}#2$}

\def\solarmass{$M_\odot$}
\def\solarmasse{M_\odot}

\def\mJyb{mJy beam$^{-1}$}

\def\tlabel#1{(\textit{#1})}

\def\cmc{cm$^{-3}$}

\def\ra#1#2#3#4{#1^\mathrm{h} #2^\mathrm{m} #3^\mathrm{s}_{^\textrm{.}} #4}
\def\dec#1#2#3#4{#1\degr #2\arcmin #3^{\prime\prime}_{^\textrm{.}}#4}

\def\mH{m_\textrm{\scriptsize H}}
\def\Ro{R_\textrm{\scriptsize 0}}

\def\To{T_\textrm{\scriptsize 0}}
\def\no{n_\textrm{\scriptsize 0}}

\def\H2{H$_2$}
\def\N2HP{N$_2$H$^+$}

\def\cCO{C$^{18}$O}

\def\NH3{NH$_3$}

\def\putfig#1#2#3{\epsfig{scale=#1,angle=#2,figure=#3}}
\def\leftblank#1{}

\newcounter{mfigure}[section]
\newenvironment{mfigure}[1][]{\refstepcounter{mfigure}\par\medskip
   \noindent \textbf{Figure~\themfigure. #1} \rmfamily}{\medskip}

\title{Spiral Structures in an Embedded Protostellar Disk Driven by Envelope
Accretion}

\author{Chin-Fei Lee$^{1,2*}$, Zhi-Yun Li$^{3}$, \& Neal J. Turner$^4$}

\begin{document}

\maketitle

\begin{affiliations}
 \item Academia Sinica Institute of Astronomy and Astrophysics, P.O. Box 23-141, Taipei 106, Taiwan
 \item Graduate Institute of Astronomy and Astrophysics, National Taiwan 
   University, No.  1, Sec.  4, Roosevelt Road, Taipei 10617, Taiwan
 \item Astronomy Department, University of Virginia, Charlottesville, VA 22904, USA
 \item Jet Propulsion Laboratory, California Institute of Technology, Pasadena, CA 91109, USA
\end{affiliations}


\begin{abstract}

Hydrodynamical simulations show that a pair of spiral arms can form in the
disk around a rapidly-growing young star and that the arms are crucial in
transporting angular momentum as the disk accretes material from the
surrounding envelope\citep{Bate1998,Rice2003,Tomida2010,Harsono2011}.  Here
we report the detection of a pair of symmetric spiral structures in a
protostellar disk, supporting the formation of spiral arms in the disk
around a forming star.  The HH 111 VLA 1 source is a young Class I source
embedded in a massive infalling protostellar envelope and is actively
accreting, driving the prominent HH 111 jet.  Previous observations showed a
ring of shock emission around the disk's outer edge\citep{Lee2016},
indicating accretion of the envelope material onto the disk at a high rate. 
Now with ALMA observations of thermal emission from dust particles, we
detect a pair of spiral arms extending from the inner region to the disk's
outer edge, similar to that seen in many
simulations\citep{Bate1998,Rice2003,Tomida2010,Harsono2011}.  Additionally,
the disk is massive, with Toomre's Q parameter near unity in the outer parts
where the spiral structures are detected, supporting the notion that
envelope accretion is driving the outer disk gravitationally unstable. 
In our observations, another source, HH 111 VLA 2, is spatially resolved for
the first time, showing a disk-like structure with a diameter of $\sim$ 26
au and an orientation nearly orthogonal to that of the HH 111 VLA 1 disk.

\end{abstract}




Spiral structures have been recently detected at (sub)millimeter wavelengths
in protoplanetary disks whose protostellar envelopes have fully or partly
dispersed\citep{Perez2016,Andrews2018,Huang2018,Kurtovic2018}.  These
systems are in the late stages of their stars' growth, when planets are
supposed to form.  The spirals in most of these systems have $m=2$ symmetry
and trail relative to the disk rotation.  The fact that systems are
seen mostly with $m=2$ structures could be due to observational bias since
higher modes are harder to resolve\citep{Hall2018}.  Unlike the spirals
detected in near-infrared and optical images that trace scattered star
light, the (sub)millimeter spirals are thermal emission from dust that has
settled closer to the midplane of the disk where the bulk of the mass
resides, allowing us to investigate how the spirals are excited.  The
spirals could be induced by stellar or substellar companions, and thus
potentially be linked to planet
formation\citep{Perez2016,Andrews2018,Huang2018,Kurtovic2018,Hall2018}.  The
spirals in some systems may be driven by global gravitational instability
(GI)\citep{Meru2017,Huang2018,Forgan2018}, in which the disks are massive
enough that their self-gravity overcomes the thermal pressure and
shear\citep{Toomre1964,Gammie2001}.  Some of the spirals may be driven by
both a companion and GI\citep{Meru2017,Forgan2018}.

These spiral structures in protoplanetary disks near the end of star
formation naturally raise the question of whether spiral structures can also
appear in the early embedded phase of star formation when gas and dust from
the surrounding molecular cloud core are still infalling onto the disk. 
Indeed, a prominent spiral structure has been detected in the younger, Class
0 system L1448 IRS3B\citep{Tobin2016}.  This is a triple system with a close
binary at the center of the disk and a tertiary source in the outer disk. 
The disk in this system appears to have recently undergone GI, inducing the
prominent spiral structure centered on the close binary with the third
source located along the spiral structure in the outer
disk\citep{Tobin2016}.  The prominent spiral structure has apparently led to
the formation of the tertiary source through disk fragmentation.  When the
disk gas has a cooling time shorter than about half the orbital
period, the cooling can drive the disk toward instability and the collapse
of one or more fragments to form new gravitationally-bound
bodies\citep{Gammie2001,Rice2003}.  Another potential example is the Class I
binary system BHB07-11, where spiral-like structures are detected in dust
continuum emission, although apparently in the inner protostellar envelope
outside the compact circumbinary disk\citep{Alves2017}.

Interestingly, many simulations have also shown that a pair of symmetric
spiral structures can appear in the main protostellar mass accretion
phase\citep{Bate1998,Rice2003,Tomida2010,Harsono2011}, starting from the
Class 0 and persisting to the end of the Class I phase\citep{Tomida2017}. 
Unlike the spiral structure detected in L1448 IRS3B\citep{Tobin2016}, these
are symmetric and play a critical role in transferring orbital angular
momentum from the inner to the outer disk, allowing accretion even in the
absence of other forms of angular momentum transport such as
magneto-rotational instability or a disk wind\cite{Turner2014}.  In this
paper, we report the detection of a pair of symmetric spiral structures in
the protostellar disk of an embedded Class I source, in support of this
possibility.

The HH 111 VLA 1 source is located in Orion at a distance of 400 pc.  With a
bolometric temperature of only 78 K\citep{Froebrich2005}, it is a Class I
source and therefore young.  It is still deeply embedded in a massive
infalling envelope\citep{Lee2010,Lee2016}, and is both actively accreting
and driving the prominent HH 111 jet that has a length of $\sim$ 6.7 pc (at
the remeasured distance)\cite{Reipurth1997}.  A rotating disk is detected
toward the center with a radius of $\sim$ 200
au\citep{Lee2010,Lee2011,Lee2016} producing polarized dust continuum
emission\citep{Lee2018BDisk}.  The central star has a mass $M_* \sim
1.5\pm0.5$\solarmass{}\citep{Lee2010,Lee2011,Lee2016}.  The infall rate in
the envelope towards the center is estimated at $\sim$ $4.2\times10^{-6}$
\solarmass{} yr$^{-1}$ $^($\citep{Lee2010}$^)$.   Assuming mass flows at
the same rate in the disk where $\sim$ 30\% of the flow is diverted into the
jet and wind\citep{Shu2000,Konigl2000}, the stellar accretion rate would be
${\dot M_*} \sim$ $2.9\times10^{-6}$ \solarmass{} yr$^{-1}$, resulting in an
accretion age $M_* / {\dot M_*} \sim$ $5\times10^5$ yrs. Another source VLA
2 is also detected at  a projected distance of $\sim$ 1200 au to the
west, forming a wide binary with the VLA 1 source.

Figure \ref{fig:cont} presents our ALMA 343 GHz continuum map towards the
two VLA sources at \arcsa{0}{04} (16 au) resolution, showing the thermal
dust emission around them.  The emission around the VLA 2 source is
spatially resolved for the first time, appearing as a disklike structure,
with a Gaussian deconvolved size (diameter) of \arcsa{0}{064} (25.6 au)
along the major axis at position angle 118.3\degree{}.  Since this disklike
structure is roughly perpendicular to another jet, HH 121, the VLA 2 source
likely drives the HH 121 jet, as suggested by observations of the CO
outflow\citep{Sewilo2017}.  Note that the VLA 2 disklike structure is
nearly orthogonal to the VLA 1 disk, indicating the two components of this
wide protobinary system are strongly misaligned.


The VLA 1 source powers the prominent HH 111 jet directed east-west,
as indicated by the red and blue arrows.  The surrounding emission shows a
spatially resolved disk perpendicular to the HH 111 jet, with its near side
tilted to the southeast.  To better see the disk structure, we rotate the
disk to align its major axis with the vertical (Fig.  \ref{fig:cont}b).  As
can be seen, the disk has a radius of $\sim$ \arcsa{0}{4} (160 au), where
the rotation was previously found to be roughly Keplerian in
\cCO{}$^($\citep{Lee2016}$^)$.  The aspect ratio of the disk structure is
$\sim$ 0.3, indicating that the disk axis is inclined by $\sim$ 18\degree{}
to the plane of the sky, closely aligned with the HH 111 jet, which has an
inclination angle of 13\degree$-$19\degree{} (after the proper motion is
updated for the new distance)\cite{Reipurth1992}.

Figure \ref{fig:cont}c shows the map deprojected by the inclination angle,
revealing a roundish compact core and a pair of faint spiral structures in
the outer disk.  In order to better see the spirals, we first made an
annularly averaged map of the continuum (see Figure \ref{fig:spiral}b) then
subtracted it from the deprojected map\citet{Kurtovic2018}.  Figure
\ref{fig:spiral}c shows the difference map, which clearly shows a pair of
symmetric spiral structures labeled ``NE" and ``SW" extending from within
$\sim$ \arcsa{0}{1} (40 au) of the central source to the outer edge of the
disk.  To guide the eye, the emission peaks of the spirals are marked with
green circles.  As discussed in Methods, the positions of these emission
peaks are obtained from Gaussian fits to the emission intensity in the
radial direction.  The spiral structures trail the rotation of the disk
measured in molecular lines\citep{Lee2016}.  Near the central source, two
barlike structures connect to the inner ends of the spiral structures. 
Since they are not clearly seen in the original map, these could be
artifacts of the deprojection of the central region, which is spatially
unresolved in the observation.




Two popular mechanisms have been proposed to account for the spiral
structures in protoplanetary disks\citep{Perez2016,Huang2018}.  One is the
global GI due to the self-gravity of the disk itself, which can produce a
pair of symmetric, logarithmic spiral arms with a constant pitch
angle\citep{Forgan2018}.  The other is a planet or a stellar
companion, which can also induce one or more spirals but with the
pitch angle increasing towards the location of the planet or
companion\citep{Bae2018,Forgan2018}.


Here we explore these two possibilities for the spiral arms in the
deeply-embedded VLA 1 protostellar disk by studying their pitch angles. 
Figure \ref{fig:fitspiral} shows the two spiral arms, NE and SW, in polar
coordinates, with $R$ the radius and $\theta$ the polar angle.  The data
points correspond to the emission peaks in Figure \ref{fig:spiral}c.  We
fitted the points with the two spiral shapes, one logarithmic with
$R=R_0\exp^{a \theta}$ and the other Archimedean with $R=R_0+ b \theta$,
where $b$ is the increase rate of the spiral radius with the angle, using
the nonlinear least-squares Marquardt-Levenberg algorithm.
Tables \ref{tab:log} and \ref{tab:Arc} list the fitting results with these
two spiral structures.  Based on the $\chi^2$ values, it appears that the SW
arm is slightly better fitted by a logarithmic shape, while the NE arm is
slightly better fitted by an Archimedean shape.  The pitch angle is $\sim$
16\degree{} for the SW arm and $\sim$ 13\degree{} for the NE arm, both of
which are similar to the pitch angle of the spirals in Elias 2-27 ($\sim
16^\circ$)\citep{Huang2018}.  Since neither shape fits both
arms best, higher-resolution and deeper observations are
needed to better distinguish these two possibilities.




Within the uncertainties of the current observations,  the spiral structures
in the VLA 1 disk are consistent with the logarithmic pattern expected from
a global GI.  The spiral structures are roughly symmetric and can be
traced down to similar distance on opposite sides of the central source,
also as expected for a global GI.  Simulations of gravitationally unstable
disks appear capable of reproducing the general morphology of the spiral
structures in the VLA 1 disk\cite{Tomida2017,Hall2018}.  For example, in the
simulations of Tomida et al.  (2017)\citep{Tomida2017}, the disk has a
central core and an outer pair of symmetric spiral arms appearing in the
outer part of the disk, resembling Figure \ref{fig:spiral}c.  Interestingly,
in their simulations the spiral structures disappear and reform every few
rotations as the disk accretes mass from the envelope and becomes unstable
again.  Such recurrent spiral arms first appear in the Class 0 phase and
persist until accretion declines towards the end of the Class I phase.
Spirals are expected to disappear soon after accretion has
stopped\citep{Hall2019}. Accretion of envelope material onto the disk may
trigger the development of spiral structures\citep{Hennebelle2016} and
push power into lower-m modes\citep{Harsono2011} even in
non-self-gravitating disks.

Such recurrent spiral arms can also appear in the VLA 1 disk because the
central source is still young and in the Class I phase.  The source is
deeply embedded in an envelope with a mass of $\sim$ 0.6 \solarmass{},
extending out to $\sim$ 7000 au from the VLA 1
source\citep{Lee2010,Lee2016}.  The envelope is spiraling in towards the
disk at a rate of $\sim 4.2\times10^{-6}$ \solarmass{}
yr$^{-1}$$^($\citep{Lee2010}$^)$.  This high infall rate is consistent with
fast accretion within the disk through GI-driven spirals that
transport angular momentum efficiently\citep{Dong2015,Hall2016}.  In
addition, a ring of envelope-disk accretion shock emission from the SO
molecule has been detected around the outer edge of the disk\citep{Lee2016},
providing strong evidence for mass accretion from the envelope onto the
disk.  Moreover, as discussed in Methods, the millimeter continuum
emission from the VLA 1 disk can be reasonably fitted with a simple flared
disk model.  The fit indicates the VLA 1 disk has a relatively high mass of
0.33-0.50 \solarmass{}, or 22\%-33\% of the protostellar mass. 
Therefore, the resulting Toomre's Q value is less than 1.5 for the outer
part of the disk with a radius $\gtrsim$ 100 au (Supplementary Figure
\ref{fig:Qvalue}), further supporting the possibility of global
non-axisymmetric GI as the driver for the spiral arms\citep{Durisen2007}. 
If GI is occurring in the massive VLA 1 disk, it must play an important role
in facilitating accretion by transporting angular momentum
outward\citep{Tomida2017}.




The spirals in the VLA 1 disk could instead be raised by a planet or
companion.  Our fit of the Archimedean form to the spirals indicates the
pitch angle increases inwards to $\sim$ 40 au (\arcsa{0}{1}), where we would
thus expect the planet or companion to be located and to clear a gap in the
disk.  However, since the disk seems to be too massive for a planet to open
a gap detectable at the current resolution\citep{Baruteau2008}, observations
at higher resolution are needed to search for a gap and check for this
scenario.  In addition, since it has been suggested that the spiral arms in
Elias 2-27 are due to a so-far undetected external companion in the disk
plane\citep{Meru2017}, it is natural to ask whether the spiral arms in the
VLA 1 disk can also be driven by an external companion.  One possibility is
VLA 2, but we believe this is not likely to be the driver because the disk
of VLA 1 is highly inclined with respect to the plane of the sky, being
nearly edge-on.  If VLA 2 lies on the disk plane of VLA 1, it would have to
be located at a distance much larger than its projected separation of $\sim
1,200$~au, which would greatly reduce its gravitational influence on the VLA
1 disk.  If VLA 2 is located far from the VLA 1 disk plane, in addition to
inducing spiral arms it should warp the disk. No sign of warping is
detected in our continuum observations; higher resolution continuum and
especially line observations are desirable to put tighter constraints on any
disk warping.  Another possibility is a putative stellar companion of
$\sim 1 M_\odot$, inferred to lie at a distance of $\sim 186$~au from VLA 1
based on the wiggling of the HH 111 jet and
counterjet\citep{Noriega-Crespo2011}.  However, such a massive young star
should have a disk detectable with our sensitive ALMA continuum observation
if it is at an evolutionary stage similar to that of VLA 1, yet
nothing is seen.  Further observations are needed to check the
possibility that a companion in a suitable position has been missed.



\newcommand\aap{{\it Astron. Astrophys.\,}}
\newcommand\apjl{{\it Astrophys. J. Lett.\,}}
\newcommand\apj{{\it Astrophys. J.\,}}
\newcommand\apjs{{\it Astrophys. J. Suppl.\,}}
\newcommand\aj{{\it Astron. J.\,}}
\newcommand\araa{{ARAA}}
\newcommand\nat{{\it Nature\,}}
\newcommand\mnras{{\it Mon. Not. R. Astron. Soc.\,}}


\noindent
{\bf Correspondence and requests for materials} should be addressed to C.-F.L.

\begin{addendum}
\item This paper makes use of the following ALMA data:
ADS/JAO.ALMA\#2015.1.00037.S and 2017.1.00044.S.  ALMA is a partnership of
ESO (representing its member states), NSF (USA) and NINS (Japan), together
with NRC (Canada), NSC and ASIAA (Taiwan), and KASI (Republic of Korea), in
cooperation with the Republic of Chile.  The Joint ALMA Observatory is
operated by ESO, AUI/NRAO and NAOJ.  C.-F.L.  acknowledges grants from the
Ministry of Science and Technology of Taiwan (MoST 107-2119-M- 001-040-MY3)
and the Academia Sinica (Investigator Award AS-IA-108-M01).  Z.-Y.L.  is
supported in part by NSF grants AST-1716259, 1815784 and 1910106 and NASA
grant 80NSSC18K1095.  N.J.T.'s contribution was carried out at the Jet
Propulsion Laboratory, California Institute of Technology, under contract
with NASA and with the support of NASA grant 17-XRP17\_2-0081.

\item[Author Contributions]

C.-F. Lee led the project, analysis, discussion, and drafted the manuscript. 
Z.-Y.  Li and N. J. Turner commented on the manuscript and participated in the
discussion. 

\item[Competing interests]
The authors declare no competing financial interests.
\end{addendum}


 
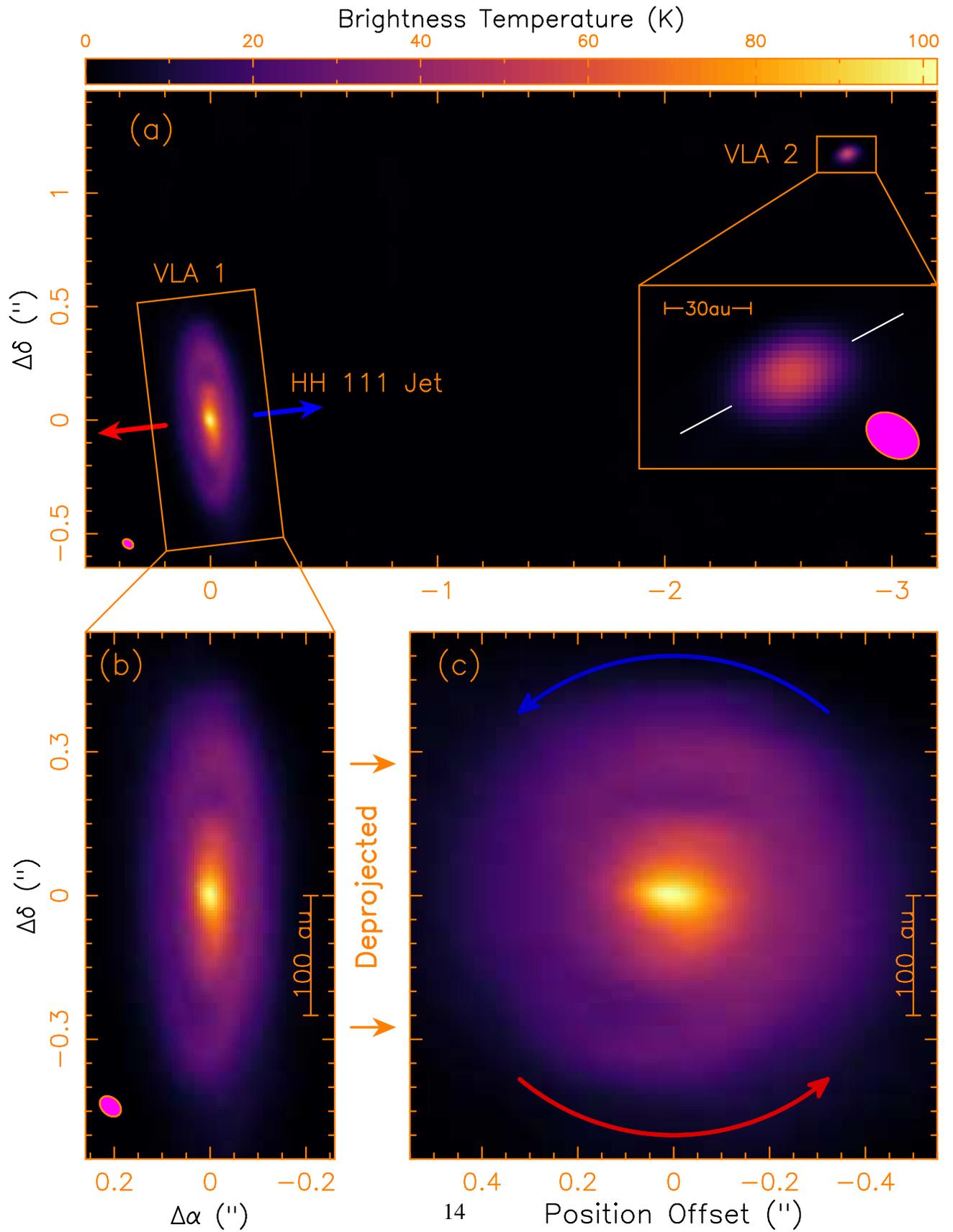
\begin{figure}
\centering
\putfig{1.2}{270}{f1.eps} 
\end{figure}
\begin{mfigure}
{
{\bf The 343 GHz continuum map towards the VLA 1 and 2 sources.}
The peak positions are ICRS $\alpha_{(2000)}=\ra{05}{51}{46}{2537}$,
$\delta_{(2000)}=\dec{+02}{48}{29}{643}$ and
$\alpha_{(2000)}=\ra{05}{51}{46}{0665}$,
$\delta_{(2000)}=\dec{+02}{48}{30}{815}$, respectively for the VLA 1 and VLA 2 sources.
The synthesized beam has a size of \arcsa{0}{050}$\times$\arcsa{0}{036}.  
Blue and red arrows in panel (a) indicate the
orientations of the blueshifted (western) and redshifted (eastern)
components of the HH 111 jet, respectively, emanating from the VLA 1 source.  
The inset is a zoom-in towards the VLA 2 source, with
white lines marking the major axis of the continuum emission.
Panel (b) shows a zoom-in on the disk around the VLA 1 source, rotated 
so the major axis is vertical.
Panel (c) shows the VLA 1 disk deprojected by the inclination angle.
Blue and red curved arrows show the direction of the disk rotation.
\label{fig:cont}} 
\end{mfigure}

\begin{figure}
\centering
\putfig{0.73}{270}{f2.eps} 
\end{figure}
\begin{mfigure}
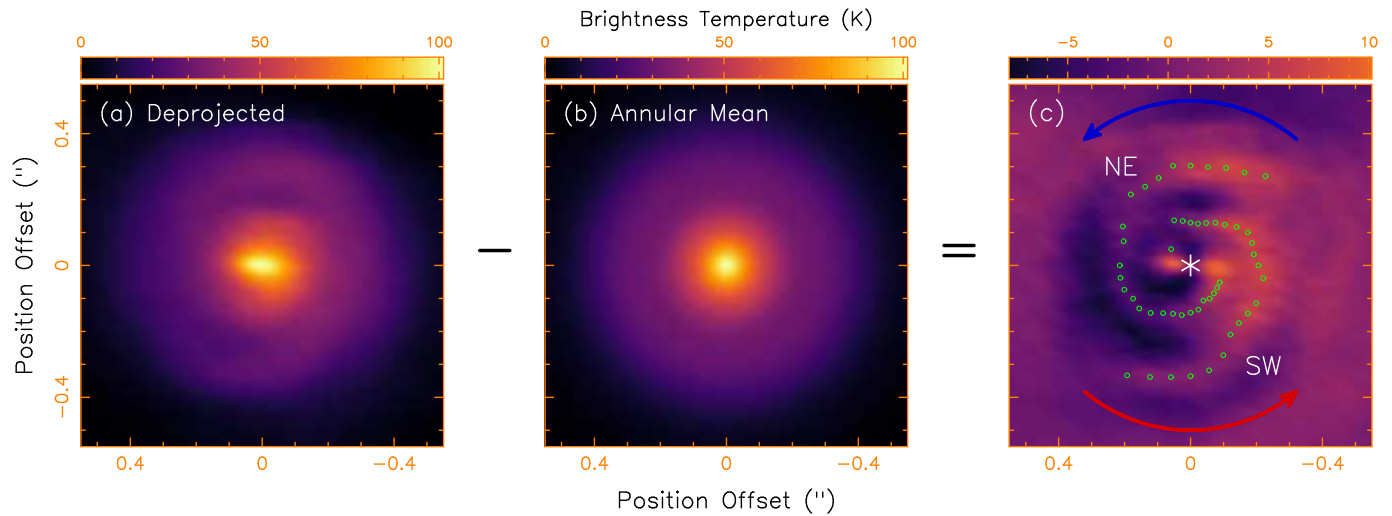
{{\bf The spiral structure in the disk around the VLA 1 source.}
Panel (a) is the same as Figure \ref{fig:cont}c
and shows the disk deprojected. In panel (b),
the structure is annularly averaged.
Panel (c) shows the map obtained by subtracting the annular mean map 
from the deprojected map, revealing the underlying spiral features NE and SW,
which have intensities $\sim$ 5-20\% of the local annular average.
Blue and red curved arrows again show the direction of the disk rotation.
Green open circles mark the emission peaks of the spirals,
outlining the spiral features.
\label{fig:spiral}}
\end{mfigure}

\begin{figure}
\centering
\putfig{0.6}{270}{f3.eps} 
\end{figure}
\begin{mfigure}
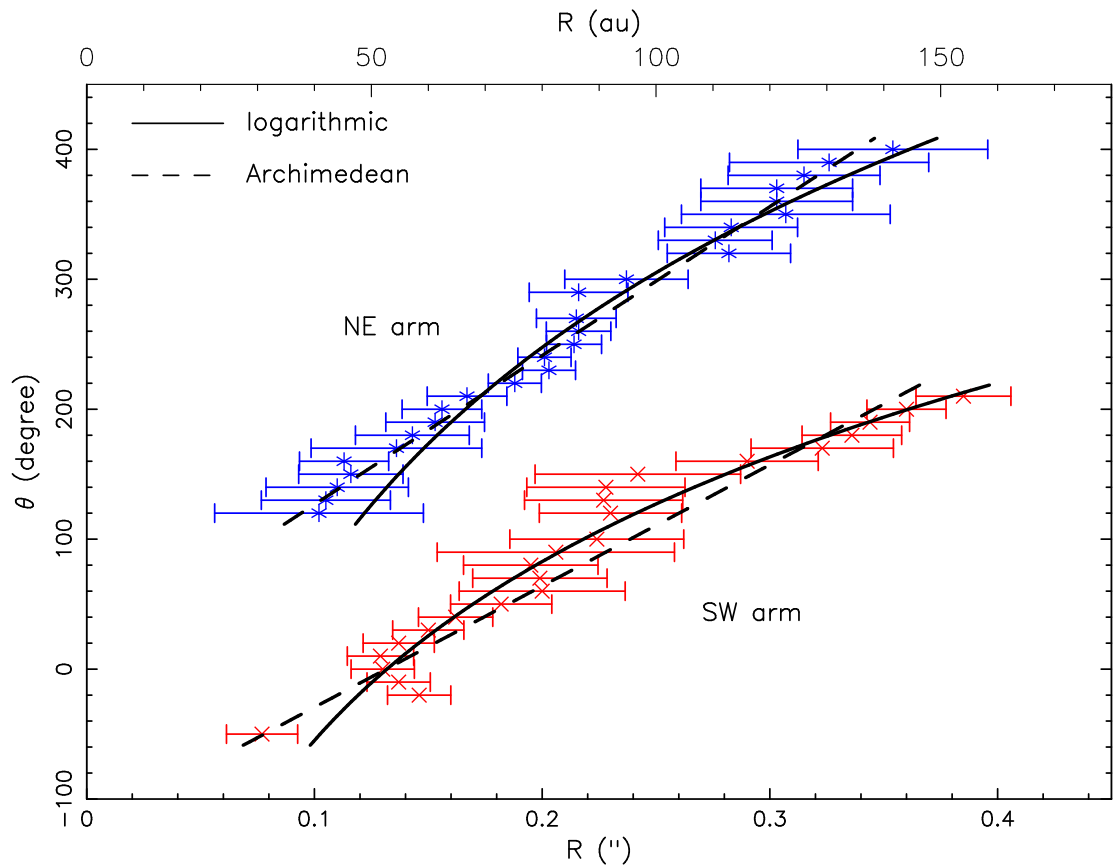
{{\bf Two fits to each of the two spiral features in the VLA 1 disk.}
The NE and SW arm are offset vertically 
for clarity.  Error bars show the arms' widths as measured by the Gaussians
that best fit the emission intensities along rays from the central source.
Each arm is fit in turn with a logarithmic (solid black curve) and an
Archimedean spiral (dashed black curve).
\label{fig:fitspiral}}
\end{mfigure}

\newcounter{mtable}[section]
\newenvironment{mtable}[1][]{\refstepcounter{mtable}\par\medskip
   \noindent \textbf{Table~\themtable. #1} \rmfamily}{\medskip}

\begin{table}
\small
\centering
\begin{mtable}
\bf Best-Fit Logarithmic Spirals\\
\label{tab:log}
\end{mtable}
\begin{tabular}{ccccc}
\hline
Arm & $R_0$       & $a$     & $\chi^2$ & Pitch Angle \\
    & (arcsecond) & (radian$^{-1}$) & & (Degree)  \\
\hline\hline
SW & \arcsa{0}{132}$\pm$\arcsa{0}{003} & 0.288$\pm$0.010 & 2.07 & 16.1\degree{}$\pm$0.6\degree{}   \\ 
NE & \arcsa{0}{077}$\pm$\arcsa{0}{004} & 0.222$\pm$0.011 & 2.88  & 12.5\degree{}$\pm$0.6\degree{} \\ 
\hline
\end{tabular}
\end{table}

\begin{table}
\small
\centering
\begin{mtable}
\bf Best-Fit Archimedean Spirals\\
\label{tab:Arc}
\end{mtable}
\begin{tabular}{ccccc}
\hline
Arm & $R_0$       & $b$     & $\chi^2$ & Pitch Angle$^\dagger$ \\
    & (arcsecond) & (arcsecond radian$^{-1}$) & & (Degree)  \\
\hline\hline
SW & \arcsa{0}{132}$\pm$\arcsa{0}{005} & 0.061$\pm$0.003 & 3.77 & 23.4\degree{}$\pm$1.2\degree{} $-$ 8.8\degree{}$\pm$0.4\degree{}   \\ 
NE & $-$\arcsa{0}{011}$\pm$\arcsa{0}{008} & 0.050$\pm$0.002 & 1.41  & 19.1\degree{}$\pm$0.8\degree{} $-$ 7.2\degree{}$\pm$0.4\degree{}  \\ 
\hline
\multicolumn{5}{l}{$\dagger$:
 Pitch Angle decreases from 60 au (\arcsa{0}{15}) to 160 au (\arcsa{0}{4})}\\
\end{tabular}
\end{table}

\clearpage

\begin{methods}
\subsection{Observations} 

We observed the HH 111 VLA 1 system with ALMA in Band 7 at $\sim$ 343 GHz in
Cycle 5 (Project ID: 2017.1.00044S).  Three observations were carried out in
C43-8 configuration using 47 antennas, two on 2017 November 28 and one on
2017 December 1, with a total time of $\sim$ 109 minutes on target.  The
projected baseline lengths were $\sim$ 60-8550 meters.  One pointing
centered at the VLA 1 source was used.  The primary beam (field of view) had
a size of $\sim$ \arcsa{17}{8}, covering both the VLA 1 and VLA 2 sources. 
The correlator was set to have four continuum windows, centered at 336.5, 338.5,
348.5, and 350.5 GHz for a total bandwidth of $\sim$ 8 GHz.  The maximum
recoverable size (MRS) was \arcsa{0}{54}.  In order to restore a larger size
scale, we combined these observations with our previous Cycle 3
observations\citep{Lee2018BDisk}, which had the same correlator setup but a
larger MRS of $\sim$ \arcsa{1}{4}.  Thus, there is no noticeable missing
flux in our maps of the disks around the two VLA sources, which have a size
$\lesssim$ \arcs{1}.


The CASA 5.1.1 package was used to calibrate the $uv$ data obtained from our
observations.  Quasar J0510+1800 was observed as a passband calibrator and a
flux calibrator, and quasar J0552+0313 (a flux of $\sim$ 0.24$\pm0.10$ Jy)
was observed as a gain calibrator.  We also performed a phase-only
self-calibration to improve the map fidelity.  As mentioned above, we
combined our observations with our previous Cycle 3 observations to
avoid any missing flux for the disks.  We used a super-uniform weighting
with a robust factor of 0.25.  The resulting synthesized beam (resolution)
has a size of \arcsa{0}{050}$\times$\arcsa{0}{036} at a position angle
(P.A.) of $\sim$ 55\degree{}.  The rms noise level is $\sim$ 0.10 \mJyb{}
($\sim$ 0.58 K).


\subsection{Spiral Structures \label{sec:spiral}}

Supplementary Figure \ref{fig:findspiral} shows the spiral structures in the
HH 111 VLA 1 disk presented in Figure \ref{fig:spiral}c.  We locate the
emission peak positions of the spirals by fitting by eye a Gaussian
to their emission intensity in the difference map cut along lines of
constant position angle every 10\degree{}.  Contours are also plotted
outlining the spiral arms.  The uncertainties in the peak positions are
shown by radial error bars, marking the width at half maximum.

\def\no{n_\mathrm{o}}
\def\na{n_\mathrm{t}}
\def\Ro{R_\mathrm{o}}
\def\Ra{R_\mathrm{t}}
\def\To{T_\mathrm{o}}
\def\Ta{T_\mathrm{t}}
\def\ho{h_\mathrm{o}}
\def\ha{h_\mathrm{t}}

\def\vk{v_\mathrm{ko}}
\def\cs{c_\mathrm{s}}
\def\vko{v_\mathrm{ko}}
\def\cso{c_\mathrm{so}}
\def\vp{v_\phi}

\subsection{Disk Model} \label{sec:model}

\def\mH2{m_{\textrm{\scriptsize H}_2}}
\def\no{n_\mathrm{o}}
\def\na{n_\mathrm{t}}
\def\Ro{R_\mathrm{o}}
\def\Ra{R_\mathrm{t}}
\def\To{T_\mathrm{o}}
\def\Ta{T_\mathrm{t}}
\def\ho{h_\mathrm{o}}
\def\ha{h_\mathrm{t}}

\def\vk{v_\mathrm{ko}}
\def\cs{c_\mathrm{s}}
\def\vko{v_\mathrm{ko}}
\def\cso{c_\mathrm{so}}
\def\vp{v_\phi}





Here we derive the properties of the VLA 1 disk in order to investigate the
origin of the spiral structures.  We adopt a parametrized flared disk model,
the same as that used to reproduce the disk emission in the
well-resolved Class 0 disk in HH 212\citep{Lee2017Disk}.  In this model, the
disk is axisymmetric, with its physical parameters specified in cylindrical
coordinates.  It has a temperature

\begin{eqnarray} 
T = \Ta (\frac{R}{\Ra})^{-q} 
\end{eqnarray} 

\noindent where $R$ is the cylindrical radius, $\Ra$ is a representative
radius to be defined below, $\Ta$ is the temperature at $\Ra$, and $q$ is
the temperature power-law index.  The model disk is in vertical
hydrostatic equilibrium, with the number density of molecular hydrogen
decreasing as a Gaussian in height above the midplane so

\begin{equation}
n= \na (\frac{R}{\Ra})^{-p} \exp(-\frac{z^2}{2 h^2}) 
\end{equation} 

\noindent where $\na$ is the number density of molecular hydrogen in the
midplane at $\Ra$, $p$ is the density power-law index, and $h$ is the
scale height.  The mass density is then $\rho=1.4 n \mH2$, 
accounting for helium at solar abundance relative to hydrogen.

In vertical hydrostatic equilibrium, the scale height $h$ comes from


\begin{equation}
\frac{h}{R} \sim \frac{\cs}{\vp} \propto \frac{R^{-q/2}}{R^{-1/2}} = 
R^{(1-q)/2}
\end{equation}

\noindent where the isothermal sound speed $\cs \propto T^{1/2} \propto
R^{-q/2}$.  The rotation speed $\vp$ is Keplerian and thus given by
$\vp=\sqrt{G M_\ast/R}$, where the protostellar mass $M_\ast \sim 1.5$
\solarmass{}, as found from the C$^{18}$O gas
kinematics\citep{Lee2010,Lee2016}.  With this scale height, increasing
monotonically in radius, we failed to fit the observed thinning of the disk
near its outer edge.  Thus, as in HH 212, we assume the disk to be
geometrically thinner near its outer edge.  This could come from
self-shielding reducing the temperature and thus the scale height in the
outermost parts\citep{Dullemond2004}.  Therefore, we define $\Ra$ to be the
radius beyond which the scale height decreases towards the outer radius of
the disk $\Ro$.  Thus, the scale height becomes

\begin{eqnarray}
h (R) \sim \ha  \left\{ \begin{array}{ll}
(\frac{R}{\Ra})^{1+(1-q)/2}  & \;\;\textrm{if}\;\; R < \Ra,  \\ 
\sqrt{1-\frac{3}{4}(\frac{R-\Ra}{\Ro-\Ra})^2} & \;\;\textrm{if}\;\; \Ra \leq R \leq \Ro
\end{array}  \right.
\label{eq:thick}
\end{eqnarray}

\noindent where $\ha\sim\frac{\cs(\Ra)}{\vp(\Ra)}\Ra$ is the scale height at
$\Ra$ and $h = \ha/2$ at $R=\Ro$.

For the dust continuum emission of the disk, we assume the dust opacity law

\begin{equation}
\kappa_\nu = 
0.1 \left( \frac{\nu}{10^{12}\textrm{Hz}} \right) ^\beta 
\;\;\textrm{cm}^2 \;\textrm{g}^{-1}
\end{equation}

\noindent which has been used before for circumstellar disks in
Taurus\citep{Beckwith1990}.  Here $\beta$ is the dust opacity spectral index
and $\kappa_\nu$ the opacity per gram of gas and dust for a gas-to-dust
mass ratio of 100. The $\beta$ value is uncertain, affecting the disk mass
needed to produce the observed total flux.  To be inclusive, we consider values
ranging from 0.6 as in the young HH 212 protostellar disk\citep{Lee2017Disk}
to 1.0 as in protoplanetary disks\citep{Beckwith1990,Andrews2009}. 
Therefore, we have $\kappa_\nu \sim 0.054 - 0.035$ cm$^2$ g$^{-1}$ at the
observing frequency of 343 GHz.

To compute the dust emission, we tilt the model disk so its axis is $\sim$
18\degree{} from the plane of the sky, as estimated from the observed aspect
ratio.  We then compute the dust emission assuming local thermodynamic
equilibrium, and integrate along a grid of lines of sight the local emission
that is attenuated by the optical depth, as done before for the HH 212
disk\citet{Lee2017Disk}.  We then convolve the resulting map with the
telescope's beam, producing a model map to be compared with the observed map
at the same resolution.


Supplementary Figure \ref{fig:model} shows our best-fit continuum map.  The
model has 6 free parameters: $\Ra$, $\Ro$, $q$, $\Ta$, $p$, and $\na$.  In
our fit, $\Ra \sim $ \arcsa{0}{4} (160 au), consistent with the observed
radius of the disk, and $\Ro \sim$ \arcsa{0}{5} (200 au) in order to produce
the faint emission in the outer edge.  The temperature power-law index
$q\sim 0.55\pm0.11$, similar to that found in protoplanetary disks in
Ophiuchus\citep{Andrews2009}.  With $\Ta \sim 68\pm13$ K, the resulting
scale height $\ha$ is approximately \arcsa{0}{068}$\pm$\arcsa{0}{007}
($\sim$ 27$\pm3$ au) at $\Ra$.  For the density, we have $p\sim 1.5\pm0.4$
and $\na \sim (6.0-9.3)\times10^{9}$ \cmc{} corresponding to the range of
opacities $\kappa\sim 0.054-0.035$ cm$^2$ g$^{-1}$.  As can be seen in
Supplementary Figure \ref{fig:model}a, the model emission contours (red) can
roughly match the observed emission contours (black with a gray-scaled
image) of the continuum.  Since the model is axisymmetric and does not
include any spiral structures, we can not fit the non-axisymmetries. 
However, since the residuals are reasonably small (Supplementary Figure
\ref{fig:model}b), the model adequately describes the radial structure.  In
addition, as can be seen in Supplementary Figure \ref{fig:model}c, since the
model disk emission is mostly optically thick along the major axis, the
temperature in our model is well constrained.  In our model, the disk has a
surface density

\begin{equation} 
\Sigma= 1.4\,\mH2 \int_{-\sqrt{2}\ha}^{\sqrt{2}\ha}
n\, dz 
\end{equation}

\noindent and a mass

\begin{equation} 
M_D= \int^{\Ro} \Sigma \,2\pi\,R\,dR \sim (0.33-0.50) \,\solarmasse 
\end{equation}

\noindent which is $\sim$ 22\%$-$33\% of the protostellar mass. Thus the disk 
is relatively massive.

Supplementary Figure \ref{fig:Qvalue} shows the model's stability against GI
as measured by Toomre's $Q$ parameter

\begin{equation} 
Q\sim \frac{\cs \Omega}{\pi G \Sigma}
\end{equation}

\noindent where $\Omega=\frac{\vp}{R}$ is the angular velocity of the disk. 
As can be seen, $Q$ is less than  1.5 for $R\gtrsim$ \arcsa{0}{25} (100
au), indicating that the disk is gravitationally unstable in its outer
half.  Since the dust opacity is uncertain, further observations are needed
to confirm the instability and check whether the inner part is also
gravitationally unstable.

If the $Q$ parameter at a given location were much less than unity, the
scale height there would be dominated by the local self-gravity rather than
the tidal gravity of the central star.  The scale height would then be
roughly\citep{Spitzer1942}:

\begin{equation} 
h_s \sim \frac{c_s^2}{\pi G \Sigma} \sim Q h 
\end{equation} 

\noindent The scale height would be significantly less than $h$.  However,
in our model, the $Q$ parameter is of order unity or larger, and the effect
of self-gravity on the scale height is moderate (at most a factor of 2).


\end{methods}

\noindent
{\bf Data availability.} This letter makes use of the following ALMA data:
ADS/JAO.  ALMA2015.1.00037.S and 2017.1.00044.S.  The data that support the
plots within this paper and other findings of this study are available from
the corresponding author upon reasonable request.

\end{document}